\documentclass[journal]{IEEEtran}

\IEEEoverridecommandlockouts
\usepackage[OT1]{fontenc} 
\usepackage{cite}
\usepackage{amsmath,amssymb,amsfonts}
\usepackage{algorithmic}
\usepackage{graphicx}
\usepackage{placeins}
\usepackage{textcomp}
\usepackage{xcolor}
\usepackage{tabu}
\usepackage{balance}

\usepackage{array}
\usepackage{subfigure}
\usepackage{multirow}

\PassOptionsToPackage{hyphens}{url}
\usepackage{hyperref}
\hypersetup{colorlinks, citecolor=green, filecolor=pink, linkcolor=red, urlcolor=blue }

\begin{document}

\title{When Services Computing Meets Blockchain: Challenges and Opportunities} 

\author{Xiaoyun Li, Zibin Zheng,~\IEEEmembership{Senior Member,~IEEE}, and Hong-Ning Dai,~\IEEEmembership{Senior Member,~IEEE}
\thanks{X. Li and Z. Zheng are with School of Data and Computer Science, Sun Yat-sen University, China and National Engineering Research Center of Digital Life, Sun Yat-sen University, Guangzhou 510006, China (email: lixy223@mail2.sysu.edu.cn and zhzibin@mail.sysu.edu.cn).}
\thanks{H.-N. Dai is with the Faculty of Information Technology, Macau University of Science and Technology, Macau (e-mail: hndai@ieee.org).}
}


\maketitle

\begin{abstract}
Services computing can offer a high-level abstraction to support diverse applications via encapsulating various computing infrastructures. Though services computing has greatly boosted the productivity of developers, it is faced with three main challenges: privacy and security risks, information silo, and pricing mechanisms and incentives. The recent advances of blockchain bring opportunities to address the challenges of services computing due to its  build-in encryption as well as digital signature schemes, decentralization feature, and intrinsic incentive mechanisms. In this paper, we present a survey to investigate the integration of blockchain with services computing. The integration of blockchain with services computing mainly exhibits merits in two aspects: i) blockchain can potentially address key challenges of services computing and ii) services computing can also promote blockchain development. In particular, we categorize the current literature of services computing based on blockchain into five types: services creation, services discovery, services recommendation, services composition, and services arbitration. Moreover, we generalize Blockchain as a Service (BaaS) architecture and summarize the representative BaaS platforms. In addition, we also outline open issues of blockchain-based services computing and BaaS. 
\end{abstract}


\section{Introduction}
\label{sec:introduction}

\IEEEPARstart{W}{e}  have witnessed the proliferation of diverse applications including finance, supply chain, public services, healthcare, all of which have been driven by the latest rapid development of information communication technology. Most of the conventional business activities can be constructed by computer software modules running on top of diverse computing facilities (from IoT devices to cloud servers). During this transformation, \emph{services computing} plays a critical role. Services computing is a computing paradigm that utilizes services as fundamental components for developing applications~\cite{papazoglou2003service}. It seeks to develop computational abstractions, architectures, techniques, and tools to support services broadly~\cite{bouguettaya2017service}. Services can encapsulate various computing infrastructures and meanwhile offer a high-level abstraction to support diverse applications. The modular services can largely improve the productivity of developers, software reusability, quality of service, and scalability of applications. Services computing covers the whole life-cycle of service provision including \textit{services creation}, \textit{services discovery}, \textit{services recommendation}, \textit{services discovery}, \textit{services composition}, \textit{services arbitration} 



\begin{figure*}[t]
    \centering
    \includegraphics[width = 0.8 \textwidth]{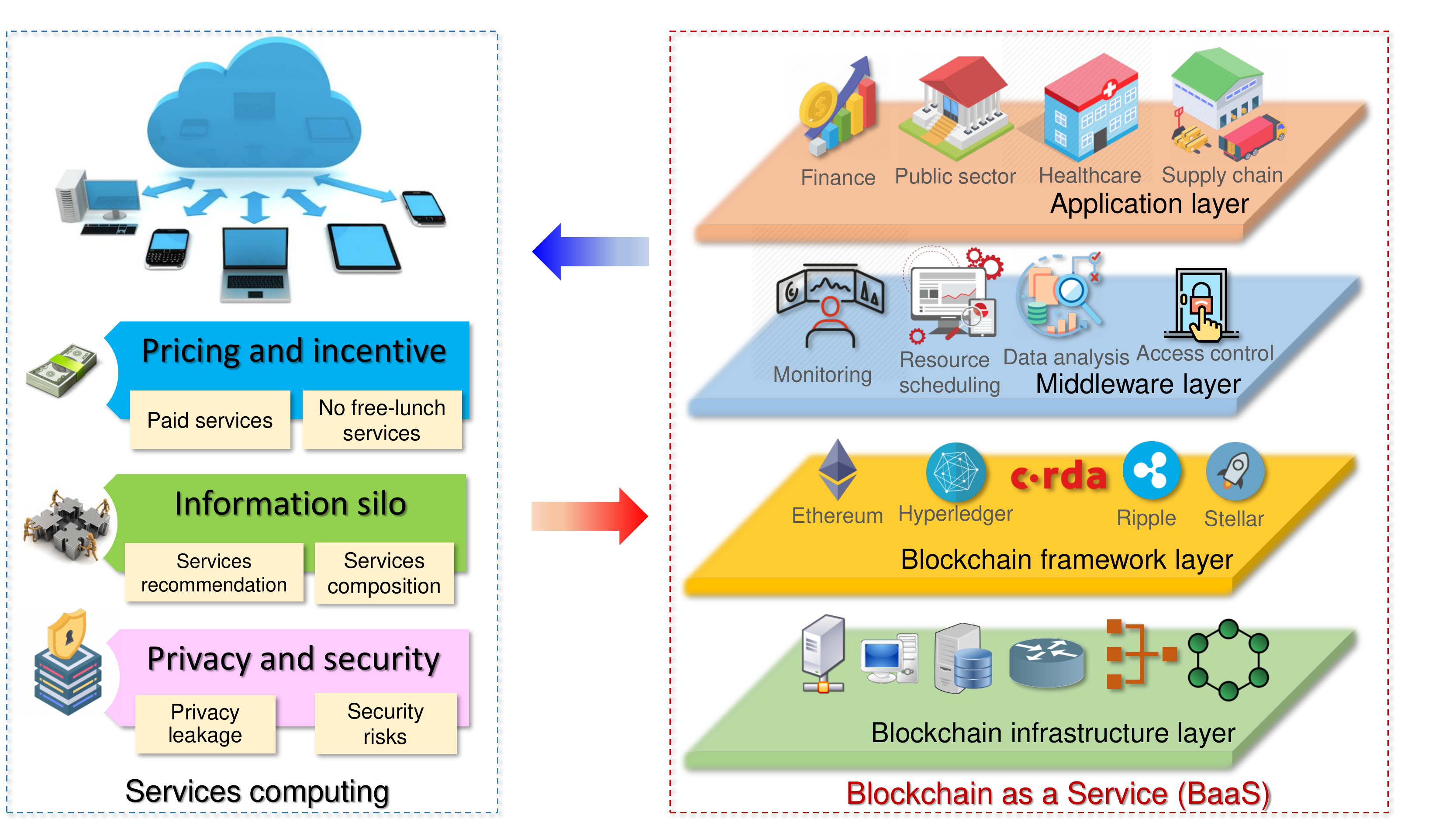}
    \caption{When services computing meets blockchain}
    \label{fig:1}
\end{figure*}

The growth of services computing has also brought a lot of challenges. These challenges can be summarized into three aspects as shown in Fig~\ref{fig:1}. We next explain them as follows:  (1) \emph{Security and privacy risks}: services vendors often collect and possess customers' privacy-sensitive data without explicit declaration~\cite{de2012cloud,goethals2007considering}. The private data can be abused or unintentionally disclosed to others without the agreement of customers. Additionally, data centers are also suffering from security vulnerabilities such as malicious attacks (like hackers or DDoS attacks) and SPFs. (2) \emph{Information Silo}: the heterogeneous information systems within an enterprise or across different business sectors have led to difficulties in information sharing and reciprocal operations among different systems, consequently forming dozens of information silos~\cite{bronsted2010service,wei2010service}. Information silo inevitably increases communication costs and lowers service quality as the recommendation of the accurate service is often based on data analysis on historical records that now have been separated in different locations (i.e., ``\emph{silos}''). (3) \emph{Pricing and incentive issues}: the pricing dilemma has hindered the development of services ecosystems ~\cite{cusumano2008changing,buyya2008market,raggett2015web}. For example, LinkedIn firstly released most of its APIs for free but had to adopt the paid APIs (instead of free APIs) when free APIs were found to be abused by selfish developers to make profits. However, the paid services may dampen the developers' enthusiasm. Moreover, the emerging application scenarios such as M2M service trading~\cite{KLiu:IoTJ19} and crowd-sourcing collaboration~\cite{JHuang:TII19} have driven the development of new pricing and incentive mechanisms.   


The advent of recent blockchain technologies brings opportunities to overcome the above challenges of services computing. Blockchain was originally designed for digital currencies such as Bitcoin~\cite{nakamoto2008bitcoin} and Ethereum~\cite{wood2014ethereum}. Thanks to recent advances in cryptography, distributed systems, consensus algorithms, and smart contract, blockchain has evolved into a trustworthy and decentralized platform to support diverse applications such as supply chain, finance, healthcare, energy, intellectual property protection, and IoT~\cite{zheng2017overview,xie2020blockchain,bashir2017mastering,WLiang:TETC20,zheng2016blockchain,hndai:blockchain-iot2019,WLiang:TII2020,Hakak:NMag20}. Blockchain can potentially solve the challenges of services computing from the following perspectives. 1) \emph{Built-in encryption and digital signature schemes} of blockchain can integrate with other security countermeasures such as authentication and access control so as to enhance the system security and preserve the data privacy. For example, data encryption and access control implemented on top of blockchain can effectively reduce the chance of data misuse and privacy leakage. 2) \emph{Decentralization} of blockchain can help to mitigate the security risks and vulnerabilities such as DDoS attacks and SPFs. Besides, the auto-execution of smart contracts can help to update the firmware so as to mitigate the security vulnerabilities \cite{christidis2016blockchains}. 3) \emph{Intrinsic incentive mechanisms} may address the pricing and incentive problem of services computing. For example, developers who contribute codes or report bugs can be rewarded a certain amount of digital currency through the automated execution of smart contracts. Therefore, integration of blockchain with services computing can overcome the challenges of services computing.

In this article, we investigate the integration of blockchain with services computing. This paper first outlines three major concerns in services computing and briefs blockchain technology. Blockchain can potentially address the challenges of services computing. The core contributions of this paper can be summarized as two aspects: 1) We investigate how blockchain can serve for services computing to address the challenges of services computing; 2) We also discuss how services computing can benefit blockchain development. In particular, we name such an alliance of blockchain and services computing as BaaS. We generalize a four-layer BaaS architecture consisting of four layers as shown in Fig.~\ref{fig:1} (from bottom to top): i) \emph{blockchain infrastructure layer}, ii) \emph{blockchain framework layer}, iii) \emph{middleware layer}, iv) \emph{application layer}. This layered BaaS architecture has the following merits including the \emph{abstraction} so as to hide the underlying complexity of diverse computing facilities and blockchain systems, \emph{general interfaces} (or services) to support diverse applications and fasten the development of services, \emph{seamless interoperability} across different underlying blockchain systems. Recently, many tech giants have announced their BaaS platforms mainly based on their incumbent cloud computing platforms. However, such BaaS platforms are still in the early stage and many open research issues such as scalability and elasticity also pose challenges on BaaS.

\begin{table}
\renewcommand\arraystretch{1.5}
\caption{Acronym Table}
\begin{center}
\begin{tabular}{l|l}
\hline
\textbf{Terms} & \textbf{Acronyms} \\
\hline \hline
Proof of Work  & PoW      \\
Proof of Stake                                                                  & PoS      \\
Delegated Proof of Stake                                                        & DPoS     \\
Single-Point Failures               & SPF      \\
Distributed Denial of Service & DDoS \\
machine-to-machine                                                              & M2M      \\
blockchain as a service                                                         & BaaS     \\
Quality-of-Service                                                              & QoS      \\
Services-Oriented Architecture                                                  & SOA      \\
Universal Description, Discovery, and Integration                               & UDDI     \\
Directed Acyclic Graph                                                          & DAG      \\
Internet of Thing                                                               & IoT      \\
Amazon Web Services                                                             & AWS     \\
IBM Blockchain Platform & IBP\\

\hline
\end{tabular}
\label{tab:acronym}
\end{center}
\end{table}
The rest of the paper is organized as follows: Section~\ref{sec:background} overviews the basic concept of blockchain and services computing. Section~\ref{sec:challenges} analyzes the challenges in current services computing. Sections~\ref{sec:sc} and~\ref{sec:bconsc} then present a survey on services computing based on blockchain and an overview on blockchain as service platforms, respectively. Finally, we conclude this paper in Section~\ref{sec:conc}. Table~\ref{tab:acronym} summarizes the key acronyms.

\section{Background}
\label{sec:background}

This section first presents an overview of services computing in Section~\ref{subsec:serv_comp} and then gives a brief introduction of blockchain in Section~\ref{subsec:blockchain}.

\subsection{Services Computing}
\label{subsec:serv_comp}




Services computing is a computing paradigm that exploits services as fundamental elements to develop applications. Services computing is composed of computational abstractions, technology architectures, and tools~\cite{papazoglou2003service,bouguettaya2017service,huhns2005service,papazoglou2007service,papazoglou2008service}. As fundamental components, services are self-describing, discoverable, reusable, and platform-independent units. They can essentially build various complex businesses. Since it is quite difficult to search the matching services, the QoS  information (e.g., reliability, availability, and response time) is the critical discriminant to distinguish among multiple similar services~\cite{zeng2004qos,ran2003model,menasce2002qos}. Meanwhile, SOA is designed to be a technological architecture model that organizes discrete services into a comprehensive application. It helps to simplify the system management process and reduce the burden of building a complicated service for enterprises. Hence, the aim of services computing is to improve the efficiency of developers, to enhance the quality of application, and to reduce the cost of building a modular software system.

\begin{figure}[t]
    \centering
    \includegraphics[width = 0.51\textwidth]{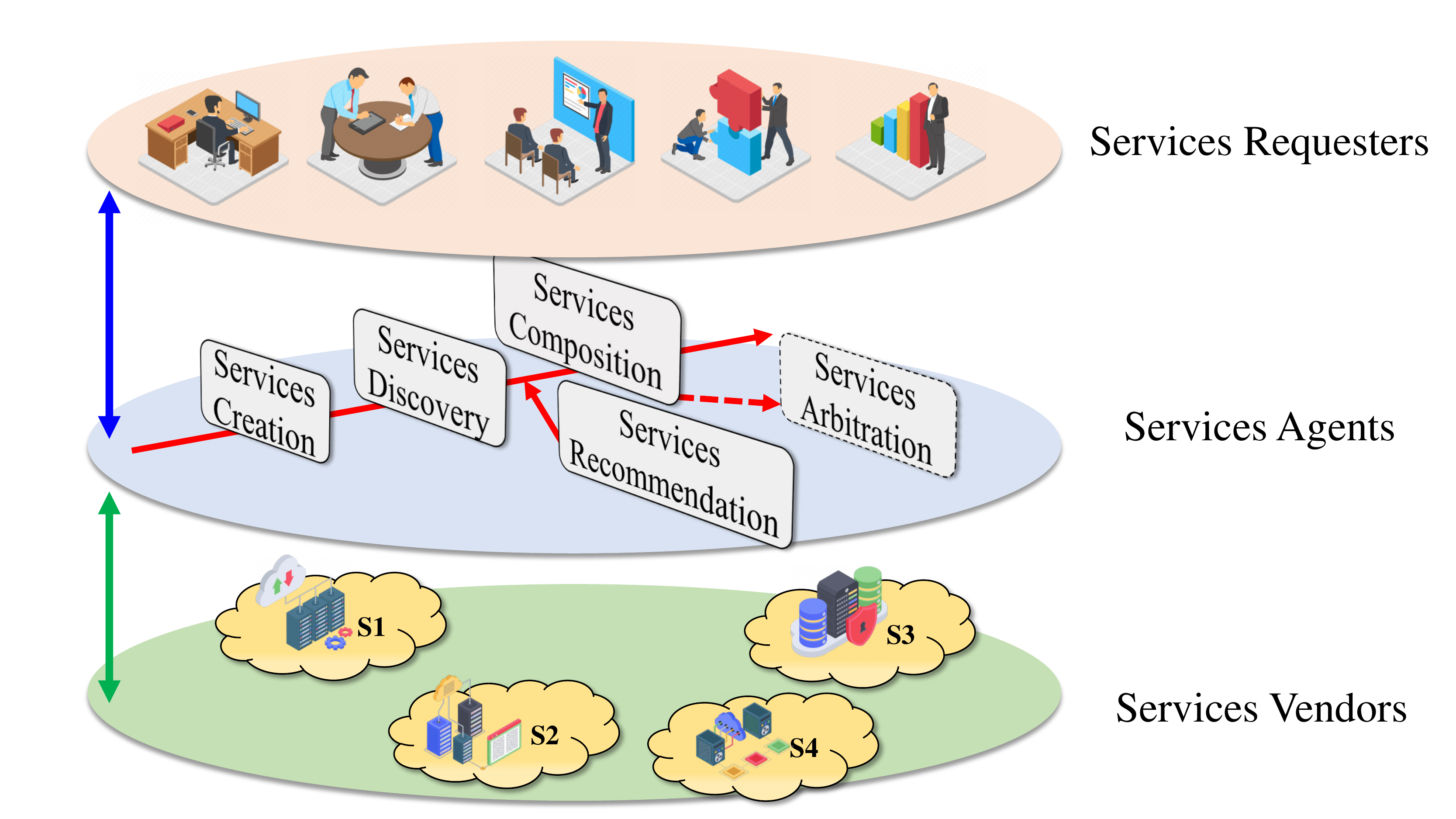}
    \caption{Services requesters, services vendors and services agents play three main roles in services computing}
    \label{fig:2}
\end{figure}



Fig.~\ref{fig:2} shows that services requesters, services vendors, and services agents play three main roles in services computing. Meanwhile, SOA can be divided into five processes: \textit{Services Creation, Services Discovery, Services Composition, Services Recommendation, Services Arbitration}. Take web services as an example. In web services, services vendors firstly publish their services with descriptions specifying their definitions and location information in the WSDL (XML-based) standard. Services requesters that indicate specific requirements can search the appropriate services in directory services enabled by UDDI which contains service creation. Services requesters then interact with the corresponding services vendors using transportation protocol such as Simple Object Access Protocol (SOAP). In addition, services agents serve as recommendation agents to recommend high-quality services using QoS-aware recommendation algorithms such as Collaborative Filtering in services recommendation and services composition~\cite{zheng2011qos,zheng2011qos,shao2007personalized}. After services discovery, composite services accessed from multiple services vendors respond to the services requesters. Although being filtered, some of the services are de facto existing faults that exert a poor influence on the system. To address this issue, service arbitration is necessary to make a justice judgment to decide the faulty party.

\begin{table*}[t]
\caption{Comparison of smart contract platforms}

\renewcommand\arraystretch{2}
\begin{center}
\begin{tabular}{c|p{5.5cm}<{\centering}p{1.5cm}<{\centering}p{2cm}<{\centering}p{1.5cm}<{\centering}p{1.5cm}<{\centering}}
\hline
\textbf{Smart Contract Platform} & \textbf{Design Purpose}& \textbf{Permission}  & \textbf{Consensus} & \textbf{Supporting Languange} & \textbf{Confidentiality}\\
\hline
\hline
{Ethereum}~\cite{ethereumwebsite} & A decentralized platform that runs smart contracts & Permissionless & PoW & Solidity & Public \\
\hline
{Hyperledger Fabric}~\cite{hyperledgerwebsite} & Mainly for enterprise and business applications & Permissioned & Pluggable & Golang, Java & Private \\
\hline
{R3 Corda}~\cite{r3website} & Corda helps business institutions to interact with each other via smart contracts & Permissioned & Raft & Kotlin, Java & Private\\
\hline
{Quorum}~\cite{quorumwebsite} & An enterprise-focused version of Ethereum & Permissioned & Raft & Golang & Private \\
\hline
{EOS}~\cite{eoswebsite} & Mainly for a decentralized system to support various decentralized applications & Permissionless & DPoS & Popular Programming Language & Public \\
\hline
{Ripple}~\cite{ripplewebsite} & A decentralized platform that sends and receives money globally without friction & Semi-Permissioned & Ripple consensus & Popular Programming Language & Private \\
\hline
{Stellar}~\cite{stellarwebsite} & Aiming at redefining financial systems to connect users, banks and payments systems & Permissionless & Stellar Consensus Protocol & Popular Programming Language & Public \\
\hline
{Neo}~\cite{neowebsite} & Digitizing assets and automating the asset management using smart contracts & Permissionless & dBFT & Popular Programming Language & Public\\
\hline
{Qtum}~\cite{qtumwebsite} & Leveraging UTXO to ensure security and supporting multiple types of virtual machines & Permissionless & PoS & X86-based Programming Language & Public\\
\hline
{Cardano}~\cite{cardanowebsite} & A decentralized public blockchain and cryptocurrency project & Permissionless & Ouroboros (a kind of PoS) & Haskell & Public\\
\hline
{IOTA}~\cite{iotawebsite} & Mainly adopting DAG technology instead of chain-like structures widely leveraged in existing blockchains & Permissionless & Tangle & Popular Programming Language & Public\\
\hline
\end{tabular}
\label{tab:scplatforms}
\end{center}
\end{table*}



\subsection{Blockchain}
\label{subsec:blockchain}
We then briefly introduce blockchain structure and smart contract, next give a comparison on popular smart contract platforms. 

\subsubsection{\textbf{Blockchain structure}}
Blockchain is an emerging technology that covers multiple computer science disciplines, such as cryptographic hash, asymmetric cryptography-based digital signature, and distributed consensus mechanisms. It is essentially a distributed ledger running in a peer-to-peer environment. Therefore, each transaction can be created, verified, and stored in a reliable network without trusted third parties. At first, a transaction is created by a peer node with the signed digital signature which is appended at the end of this transaction record. Thereafter, completed transactions are put into the transaction pool maintained by each peer node. Once the amount of transactions has reached the preset number, miners will package them into a block and propagate this event across the entire blockchain network. Then most miners compete for the right to record this block in the public ledger using a specific consensus mechanism. Finally, all the nodes reach a consensus on the public ledger and synchronize new block information to ensure all the nodes maintain the same public ledger. Iteratively, miners package their blocks following this block, consequently forming a blockchain. 

Blockchain has three key components as described as follows: 
\paragraph*{\textbf{Transaction}} Actions taken on the ledger cause changing the ledger state, e.g., concatenating a transfer record. In Bitcoin blockchain, every transaction can be verified by the signature of the previous transaction and the public key of the next owner. Once the transaction is propagated in the blockchain network, each node in the network checks its validity and verifies whether it has been unhandled. If so, this transaction will be tagged as the unconfirmed valid transaction and be propagated in the entire blockchain network.
\paragraph*{\textbf{Block}} Each blockchain contains a block header and a block body. The block header comprises Version, Previous hash, Merkle Root, Timestamp, Nonce, and Difficulty, which can be explored on the blockchain website\footnote{\url{http://blockchain.com}}. The block body includes the detailed information of transactions.

\paragraph*{\textbf{Chain}} In a blockchain network, each block is identified by a hash value that maintains a previous hash which refers to their parent block (there is an exception that the first block called the genesis block points to zero). Blocks are linked one by one in a list, consequently forming a blockchain. The chain also records all the state changes on the block thereby the source of each transaction can be traced through traversing the whole chain. 

\subsubsection{\textbf{Smart Contract}}

Smart Contract was firstly proposed by Szabo~\cite{szabo1996smart} in 1996. It is an event-driven promise written in programming languages. Smart contract is a revolution because various kinds of contractual clauses can be embedded in software, hardware, or smart agent. Once the preset conditions are met, smart contracts will run automatically in the predefined way. Therefore, electronic data can be exchanged automatically under a safe, and distributed environment. Furthermore, blockchain records every operation of smart contracts so that users can trace the actions on smart contracts. In short, a smart contract possesses four key properties: \textit{observability, verifiability, privity, enforceability}.

Ethereum firstly adopts the concept of smart contracts on blockchain~\cite{buterin2014next,ethereumwebsite}. It is a public permissionless distributed platform where users can create transactions anonymously. In contrast to the Bitcoin network system, Ethereum provides the Turing-complete programming language Solidity to develop smart contracts that are compiled down to Ethereum Virtual Machine bytecode and deployed on the Ethereum blockchain for execution. It uses the PoW consensus algorithm.

Catering for the demands from enterprises,  Hyperledger Fabric~\cite{hyperledgerwebsite} is a permissoned distributed ledger platform running on the private network where ``chaincode'' is called instead of ``smart contract''. For enterprise use, Hyperledger Fabric is designed to provide pluggable implementation delivering high confidentiality, resilience, and scalability. Additionally, Hyperledger also adopts pluggable consensus algorithms to meet business requirements. 

Similar to Hyperledger Fabric, Corda Enterprise~\cite{r3website} is a commercial version of Corda (which is originally open source) to fulfill the rising business demands. Each peer in the Corda network can only see a subset of facts on the ledger. To address the issue of non-deterministic contract execution, a service called Oracle that is used to sign the transaction if the included fact is true. It adopts the Notary services to select a consensus algorithm based on their requirements.

In a nutshell, three mainstream smart contract platforms are introduced above. Additionally, with the advent of smart contract, an increasing number of smart contract platforms have appeared. We survey and make a comparison of 11 smart contract platforms in terms of design purpose, permission, consensus, supporting language, and confidentiality, as shown in Table \ref{tab:scplatforms}. 

Many institutions or companies have developed different smart contract platforms according to different scenarios, as shown in the column ``Design Purpose'' in Table~\ref{tab:scplatforms}. Moreover, each smart contract platform has different settings with respect to permission settings, consensus schemes, supporting languages, and confidentiality due to different design purposes. As shown in Table~\ref{tab:scplatforms}, most smart contract platforms have been established on permissionless blockchains, which require no permission to join and are essentially publicly available. In contrast, permissioned blockchains strictly require an authority to access and participate in blockchains. The most representative permissioned smart contract platforms include Hyperledger Fabric, R3 Corda, and Quorum (i.e., Ethereum for enterprise). Semi-permissioned blockchains sit between permissioned and permissionless blockchains while semi-permissioned blockchains are similar to permissionless blockchains in confidentiality. Thus, permissioned and semi-permissioned blockchains are categorized as private blockchains. In Table~\ref{tab:scplatforms}, we only choose Ripple as the semi-permissioned smart contract platform. With respect to supporting language, most smart contract platforms support popular programming languages, which offer better flexibility. Ethereum especially develops a smart contract language Solidity whereas R3 Corda and Cardano adopt Kotlin and Haskell as programming languages, respectively. Regarding consensus schemes, Section~\ref{subsec:consensus} gives more details.

\subsubsection{\textbf{Consensus}}
\label{subsec:consensus}
Consensus algorithms are proposed to ensure all nodes to reach consensus in the distributed asynchronous network. Distributed consensus is essentially modelled by the Byzantine General problem, in which only one node is selected to validate the transaction without controversy~\cite{mingxiao2017review,dinh2018untangling}. To address the Byzantine General problem, there are typically two ways shown as follows.

One type is computing competence. For example, PoW~\cite{nakamoto2008bitcoin} selects the node which is the first one to randomly compute the exact hash value. In contrast to computing-intensive PoW, PoS selects the validate node according to the wealth and age (i.e., stake). In other words, the stakeholders with larger stake values can have a higher chance to verify the block. Moreover, DPoS inserts a democratic layer with the representative node to avoid the centralization in PoS.

Another type is the communication competence. Based on Byzantine Fault Tolerance (BFT), Practical Byzantine Fault Tolerance (PBFT)~\cite{castro1999practical} is a state-machine replication, in which each replica maintains the condition of services and implements the operation of services. In PBFT, the system is able to handle 1/3 faulty replicas thereby having high scalability and reliability. Paxos~\cite{lamport2001paxos} and its optimized version Raft~\cite{ongaro2014search} implement crash-tolerant state machine replications. In contrast to Raft, Ripple consensus~\cite{schwartz2014ripple} can support a large scale network via aggregating multiple sub-networks (which are trusted). Stellar Consensus Protocol~\cite{mazieres2015stellar} is a federated Byzantine agreement where the consensus first reaches each federation and is then propagated to the rest of the network. Additionally, Tangle~\cite{popov2016tangle} proposed the IOTA organization to pack the blocks into DAG instead of generating chains. In this way, it can largely reduce the cost and improve the efficiency of transactions.

\section{Challenges in Services Computing}
\label{sec:challenges}
Small (or micro) services can construct software systems and their components, thus both reconstruction and customization of software have been greatly improved. As a result, services computing has received extensive attention while it also poses a large body of challenges. In this section, we identify three major challenges in services computing: 1) security and privacy risks (in Section~\ref{subsec:security}), 2) information silo (in Section~\ref{subsec:silo}) and 3) pricing mechanisms and incentives (in Section~\ref{subsec:pricing}). 


\subsection{Security and Privacy Risks}
\label{subsec:security}


With the advent of information communication technology, people enjoy the benefits of a wide diversity of services.  According to Nielsen's Total Audience Report~\cite{electronic}, Americans aged 18 or above spend roughly ten and a half hours a day using services in different media such as mobile devices, PCs, and other electronic devices. When using these services, every action taken on the electronic devices generate personal data and logs including user behaviors, services trace logs, and user profiles. Services vendors collect these data and offer them to third parties without the consent of the data subject. As a result, the overwhelming data generated every day are at a high risk of privacy leakage and security vulnerability. We next discuss these two concerns as follows.

\subsubsection{\textbf{Privacy Leakage}}
Services computing is faced with risks of privacy leakage. Take an online bookstore as an example. Alice is a mother and she wants to buy books for her children via an online bookstore. Alice registered an account in the bookstore with her privacy-sensitive information, including name, gender, phone number, and personal address. When Alice is surfing in the bookstore, the bookstore will save all browsing records. All of these customer records will be stored in centralized servers owned by services vendors (e.g., the bookstore). These privacy-sensitive data may be misused by vendors (e.g., selling to third parties to analyze user portraits), leading to privacy leakage.



\subsubsection{\textbf{Security Risks}}

In addition to the privacy leakage, the customer privacy-sensitive data stored at centralized servers of services vendors may be hacked by malicious attackers~\cite{WLiang:IDS2020}. Moreover, when a service is invoked, it will generate running logs including system information, transaction information, faults, and errors. The system logs can help to restore systems after faults, or locate root causes when an error occurs. Generally, services trace logs can only be utilized by system developers or administrators. However, the centralization of services computing can also lead to the vulnerability of systems to either SPF or malicious attacks (e.g., DDoS attacks)~\cite{WLiang:IoTJ2020}.

Nowadays, user privacy-sensitive information (i.e., location information, preference, and political attitude) has been excessively collected and leveraged by services vendors. However, the customer data that is essentially generated by customers has been controlled by services vendors via user incidental authentication~\cite{de2012cloud}. The advent of blockchain has the potentials to address the privacy and security issues via its intrinsic encryption, pseudonymity, and decentralization. Thus, embracing blockchain technology, services computing may overcome the above challenges.

\subsection{Information Silo}
\label{subsec:silo}
An information silo is an isolated information system that is incapable of communicating with other information systems. Information silo obstructs information sharing, consequently hindering the development of data analytics. The isolated services agents divide the whole datasets into small subsets, each of which is separated from each other, consequently resulting in the information silo. We then discuss two kinds of information silo in detail as follows.



\subsubsection{\textbf{Services Recommendation}} Among the identical or similar services, \textit{services recommendation} has been proposed to recommend the appropriate services according to the predefined rules via a specific algorithm. Recently, QoS-aware recommendation algorithms~\cite{zheng2011qos,zheng2009wsrec} have been adopted in most web services recommendation systems. Since the services vendors may not guarantee the QoS (as it declares) and some QoS metrics heavily depend on the network conditions (sometimes are also related to locations of customers), it is necessary for a centralized service agent to collect data, execute algorithms and present the optimal selection results. In this case, different data collection processes occur across different systems, thereby resulting in the appearance of information silo, which leads to a low-accuracy selection and recommendation. 


\subsubsection{\textbf{Services Composition}} In services computing, \textit{services composition} encompasses the processes of creating composite services from existing services. The seamless and dynamic integration of business sectors can determine the quality of composite services. Most of the services composition architectures only collect data and execute the composite result once so that they may not share the results with others~\cite{lemos2016web}. It may require the repetition of executing a similar process again to achieve the same (or similar) service composition. Consequently, it will cause the resource waste due to the information silo.

In brief, information silo greatly reduces the accuracy of services recommendation and services composition. Also, information silo produces unnecessary duplicate costs. The isolated collecting datasets hamper the progress of services computing.  Hence, it is crucial to address information silo in services computing.

\subsection{Pricing Mechanisms and Incentives}
\label{subsec:pricing}
\begin{figure}[t]
    \centering
    \subfigure[Paid services]{\label{fig:charging_services}\includegraphics[width = 0.45\textwidth]{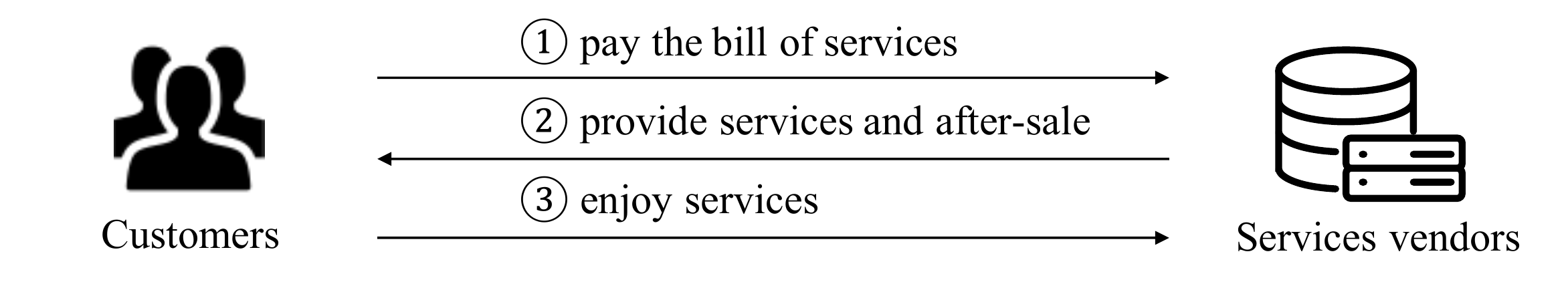}    }
    \subfigure[No free-lunch services]{\label{fig:no_free_services}\includegraphics[width = 0.45\textwidth]{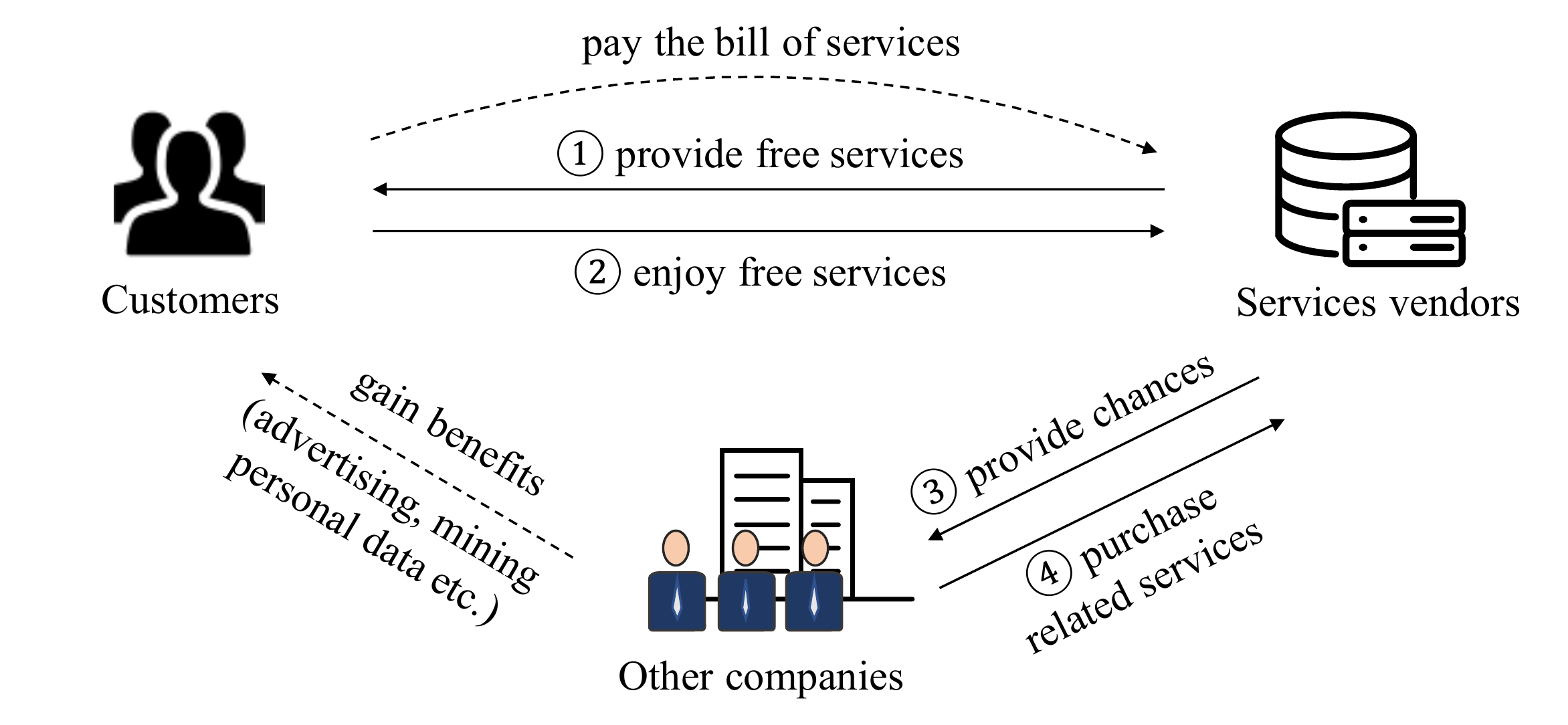}}     
    \caption{Two types of business models for services incentives}
    \label{fig:business_model}
\end{figure}

The most important driving effort for enterprises is gaining profits. Over past decades, companies have evolved from selling pure products (e.g., personal computers, laptops, and mobile phones) to service-product continua. Different from physical goods that have been possessed by buyers after purchasing, services that are impalpable can only be non-permanently owned by customers for a certain period. The purchasing process of a service is essentially a transaction made between a buyer and a seller whereas there is no physical product involved in the whole process. With the proliferation of services computing, plenty of business models for services selling spring out. We then introduce two types of business models~\cite{cusumano2008changing}: \textit{no free-lunch services} and \textit{paid services}. Fig. \ref{fig:business_model} depicts two business models in terms of transaction flows with numbered steps, in which a direct transaction is represented by a solid line while a dotted line indicates an indirect interaction. 

\subsubsection{\textbf{Paid Services}} A conventional pricing model in services computing is paid services, in which customers pay for the services directly. Once consumers pay the bill, they can enjoy the services and after-sale services provided by services vendors (under the given terms). Fig. \ref{fig:charging_services} illustrates the interactions between customers and services vendors in paid services. Pricing is a crucial issue in paid services since services vendors still need to make profits under daily computing and network operational costs after selling services while customers may mainly concern with the price-performance ratio of services. In particular, paid services allow users to enjoy services whose QoS is highly related to the price (i.e., the higher price the better QoS). However, too high prices of services may dampen customers' willingness and degrade the proliferation of services. On the other hand, a relatively low price leads to the poor QoS to customers. How to define the appropriate prices of services to balance the profits and the market has been a hot topic in services computing whereas it is challenging to design and implement an effective pricing model in practice.

\subsubsection{\textbf{No Free-lunch Services}} Fig. \ref{fig:no_free_services} shows a transaction flow in no free-lunch services. Recently, services vendors like Google, Microsoft, and Amazon provide customers with ``\emph{free}'' services such as emails, search engines, storage services, and online media streaming services. These ``\emph{free}'' services do not require customers to purchase them explicitly while they are not really ``\emph{free}''. Services vendors can essentially make profits through the following manners. The first manner is explicit advertisements. For example, customers are reluctant to accept a multitude of advertisements when they enjoy ``free'' services. Another manner is implicitly leveraging and mining the customers' personal data to third parties. Services vendors collect customer data such as searching history, preference, and location information when customers are enjoying ``free'' services. Therefore, free services are essentially \emph{not free} in the cost of explicit advertising and selling personal private data to third parties. Although the general usage of ``\emph{No free-lunch}'' business model has proved its success in many Internet service providers, the privacy vulnerability (or privacy leakage) of such a model is also a double-sided sword to customers.

\textbf{Discussion}. Aside from current pricing mechanism issues, the way to design pricing and incentive mechanisms in future services computing is also challenging. In particular, the proliferation of IoT poses challenges in services computing. It is predicated by Gartner~\cite{GartnerIoT} that there will be 25 billion IoT devices by 2021. Current services computing models or components should be tailored to be IoT-oriented services in the future. During this procedure, M2M interaction will become common in IoT. It essentially requires an autonomous, automated, accurate, and efficient checkout system to support the services trading between any two IoT devices. In a nutshell, defining an appropriate pricing mechanism is one of the current challenges in services computing. However, the business models mentioned above are not sufficient to address future challenges like M2M orchestrations in IoT scenarios. Hence, we need to define an apposite pricing mechanism for IoT with the provision of IoT-oriented services. In order to establish a cooperative environment to bloom services computing markets, it is necessary to design suitable incentive mechanisms accompanying appropriate pricing mechanisms so as to facilitate services sharing.

\section{Services Computing Based on Blockchain}
\label{sec:sc}
Recently, quite a number of solutions have been proposed to address the above issues in services computing~\cite{meng2014kasr}. For example, a Service Delivery Platform (SDP) has a provision of interfaces to enable various applications released by third-party application developers~\cite{OhnishiYKMHS07,CallawayDVR10}. SDP can help to solve the information silo issue. Moreover, Buyya et al.~\cite{buyya2008market} presented a market-oriented cloud architecture to trade services. However, each of the existing solutions only focused on a specific issue. Services computing requires a more consolidated and holistic framework with superior performance and scalability to address the challenges such as \textit{pricing mechanisms and incentives, information silo, security, and privacy risks}. Blockchain that is characterized by decentralization, persistency, anonymous, and auditability can potentially solve the challenges of services computing.
We discuss how blockchain can address three challenges in services computing as follows.

\paragraph{Pricing Mechanisms and Incentives} Cryptocurrency helps to address the pricing mechanisms and incentives challenge in services computing. Doing transactions with cryptocurrency, services trading can be easily run in an autonomous, automated, efficient, and secure environment without complicated procedures of trusted third-party. Services vendors set predefined conditions in smart contracts to count services invocation time and usage frequency. Consequently, it is convenient for both services vendors and customers to make transactions on blockchain. Furthermore, the currency circulation in blockchain greatly stimulates the services trading markets.

\paragraph{Information Silo} The decentralization feature of blockchain can help to solve the information silo problem in services computing. Since all nodes are running in a distributed peer-to-peer blockchain network, they maintain the same ledger which stores service requester data, service descriptions and running logs. In this manner, geographically-isolated data centers now can synchronize all actions at the same ledger. Consequently, it is easier to collect data stored on the public ledger from most of the services vendors and services requesters.

\paragraph{Security and Privacy Risks} Owning to decentralization, persistency, anonymous, and auditability, blockchain can mitigate the security and privacy risks in services computing. In the blockchain network, each user generates transactions which can be verified by miners. Once a user sends service requests and receives the response from services vendors, the whole process will be recorded on the ledger. Therefore, any operations are traceable and auditable. Moreover, all of these processes can only be accessed by authorized users in permissioned and semi-permissioned blockchain (i.e., consortium blockchain). Therefore, without the consent of data owners, service vendors are not able to access these personal data. In this way, potential privacy violation risks can be mitigated.

There are dozens of studies on blockchain-based services computing. We categorize them into five types according to five processes: \textit{Services Creation, Services Discovery, Services Composition, Services Recommendation, Services Arbitration} of SOA as mentioned in Section~\ref{subsec:serv_comp}. Table~\ref{tab:literature} summarizes these studies. Next, we present a detailed discussion of each type as follows. 


\begin{table}
\renewcommand\arraystretch{1.5}
\caption{Content matrix of the reviewed literature}
\begin{center}
\begin{tabular}{m{4.5cm}|m{3.5cm}}

\hline
\textbf{Services Computing Components}                       & \textbf{References}                                                                    \\ \hline \hline
\multirow{2}{*}{Services Creation}       & Ruta et al.~\cite{ruta2017semantic}                     \\
                                         & Herbaut et al.~\cite{herbaut2017model}                  \\ \hline
\multirow{4}{*}{Services Discovery}      & Daza et al.~\cite{ruta2017semantic}                     \\
                                         & Ruta et al.~\cite{ruta2017semantic}                     \\
                                         & Manevich et al.~\cite{manevich2018service}              \\
                                         & Herbaut et al.~\cite{herbaut2017model}                  \\ \hline
\multirow{3}{*}{Services Recommendation} & Cai et al.~\cite{cai2019personalized}                   \\
                                         & Li et al.~\cite{li2019blockchain}                       \\
                                         & Herbaut et al.~\cite{herbaut2017model}                  \\ \hline
\multirow{4}{*}{Services Composition}    & Viriyasitavat et al.~\cite{viriyasitavat2018blockchain} \\
                                         & Carminati et al.~\cite{carminati2018confidential}       \\
                                         & Weber et al.~\cite{weber2016untrusted}                  \\
                                         & Wang et al.~\cite{wang2018smart}                        \\ \hline
\multirow{3}{*}{Services Arbitration}    & Zou et al.~\cite{zou2018proof}                          \\
                                         & Cai et al.~\cite{cai2019personalized}                   \\
                                         & Xiong et al.~\cite{xiong2019smart} \\          
\hline
\end{tabular}
\label{tab:literature}
\end{center}
\end{table}

\subsection{Services Creation Based on Blockchain}
In the past, services vendors publish their services in UDDI with XML messages. As mentioned in Section~\ref{subsec:serv_comp}, services creation plays a critical role in services trading. Therefore, blockchain-based services creation heavily depends on the holistic model of blockchain-based services trading process. 

We summarize three kinds of blockchain-based services creation as follows. In particular, Ruta et al.~\cite{ruta2017semantic} proposed a semantic-based resources registration method. It allows the co-existence of different resource domains at the same blockchain. Every resource domain is associated with different ontologies. Each ontology can be identified with a unique Uniform Resource Identifier (URI). The registration transaction only records the resource URI while the annotation reference is stored in the private node of services vendors. Zou et al.~\cite{zou2016dispute} proposed a distributed cloud services contract management scheme that allows services contracts to register at a local service registry rather than a centralized registry. Herbaut et al.~\cite{herbaut2017model} proposed a blockchain-based collaborative video delivery model, in which three blockchains are envisioned to implement the content distribution. Content providers create smart contracts for their services on provisioning blockchain.

Different from conventional services creation, no standard is established in blockchain-based services creation. But all of the blockchain-based services creation models (as mentioned above) mainly generate a service via the transaction stored on blockchain. Then each vendor can be traced through transaction records. Furthermore, services published in blockchain are still associated with text descriptions. In this way, many provisioned services are kept in all nodes throughout the whole blockchain rather than the centralized registry node so as to avoid SPF and other vulnerabilities.  

\subsection{Services Discovery Based on Blockchain}
As computational resources are widely deployed, how to locate a specific service out of the overwhelming accessible services becomes a significant problem. Services discovery is therefore to find the services or devices among the complex network environment. In the past, services vendors publish their services in directory service in UDDI with XML messages to advertise available services, while clients execute their queries via UDDI in XML messages encoded with factors like cost, performance, location, and functionality~\cite{hoschek2002web,czerwinski1999architecture}. Previous services discovery architectures rely on certificated authority to ensure a secure running environment. 


There are several blockchain-based services discovery approaches. Daza et al.~\cite{daza2017connect} proposed a discovery approach in IoT. In this way, we scan the environment by sending a hello message. Once obtaining the response messages involving blockchain peer addresses, we can search blockchains in the cloud to identify the latest activities associated with those addresses. Ruta et al.~\cite{ruta2017semantic} proposed a gossip-based approach to propagate discovery requests and aggregate results. The requester randomly selects $n$ nodes and sends a multicast request specifying semantic annotation. Nodes that receive the request then execute semantic matching of their own resources, generating a relevance-ranked list. These nodes next randomly select $m$ nodes and forward requests. Finally, the matching service returns back after following the same route. 

In addition to basic service discovery approaches based on blockchain, IBM Research Lab~\cite{manevich2018service} pointed out that there is a gap between the two tiers that hinder the rapid adoption of changes in the chaincode and endorsement policies within the client SDK. In this work, services discovery provides APIs allowing dynamic discovery of the configuration required for the client SDK to interact with the platform, alleviating the client from the huge maintenance burden.


\subsection{Services Composition Based on Blockchain}





Applications consisting of composite services can achieve good scalability and better performance. Hence, services composition can greatly promote composite services for diverse applications to lower the level of services granularity, extend application scalability, and enhance the reliability. Before blockchain was used to build services composition, several studies have been focused on peer-to-peer provisioning. Both Gu et al.~\cite{gu2004spidernet} and Benatallah et al.~\cite{benatallah2002declarative} proposed peer-to-peer paradigms that declaratively integrate existing services into composite services through pre-defined rules in a dynamic environment. Meanwhile, other studies like Xiao et al.~\cite{xiao2005qos} and Zeng et al.~\cite{zeng2004qos} leveraged QoS to combine services in order to maximize user satisfaction under constraints given by the users and the structures of the composite services. 

Recently, blockchain-based services composition frameworks can well address the centralization and trustless problems in services composition. In particular, Business Process Management (BPM) is one of the successful cases in services composition. The main object of BPM is to optimize various business procedures in order to achieve shorter latency, better QoS to end-users, and higher financial gains. The critical issues in BPM include evaluating and verifying the trustworthiness of digitized assets and transforming them accordingly. Owning to the merits of BPM, Viriyasitavat et al.~\cite{viriyasitavat2018blockchain}, Carminati et al.~\cite{carminati2018confidential} and Weber et al.~\cite{weber2016untrusted} proposed  blockchain-based frameworks for services composition on top of BPM. These frameworks can address the fundamental trust problem in collaborative process execution using blockchain. It is worth mentioning that smart contracts were adopted to automate the workflows of BPM so as to provide the transparent inter-operations of service vendors in \cite{viriyasitavat2018blockchain}.

Despite building a simple distributed framework for services composition, other researchers construct blockchain-based \emph{QoS-aware} services composition frameworks. In particular, both Wang et al.~\cite{wang2018smart} and Viriyasitavat et al.~\cite{viriyasitavat2018blockchain} proposed  QoS-aware services composition models. To optimize the runtime performance of services composition, a smart contract-based negotiation framework was proposed in \cite{wang2018smart} where the transactions are performed automatically and reliably. This can be achieved through the signed agreement between the services requesters and vendors. At runtime, if a services vendor is getting troubles, this framework can identify it and find another services vendor to replace it. Rather than using centralized services brokers, blockchain-based services composition approaches generally support automatic composition according to the runtime QoS. Smart contracts record the state of services and take down all the compositions in the transaction. Services composition thereby is more reliable and can quickly respond.

\subsection{Services Recommendation Based on Blockchain}


Some identical or similar functional services have been provisioned in services market. Therefore, services recommendation can help  to select high-quality services from a great diversity of services. Relying on the trusted third-party, existing service recommendation systems collect data and deliver recommended results while also leading to the vulnerabilities of privacy and security due to the centralization. To help users discover more potential high-quality services, services vendors need to eliminate information silo via sharing data in a secure, privacy-protected manner. To address these challenges, a fair amount of solutions have been proposed on P2P systems~\cite{canny2002collaborative,kaleli2010p2p} or secure multi-party protocols~\cite{shmueli2017secure} over the past decades. 

However, the existing distributed solutions can solely address one of the challenges in services computing and still require a trusted third-party. As a distributed system and  a public ledger, blockchain encrypts the stored data so as to protect data privacy and security with high reliability. Therefore, the above peer-to-peer schemes and multi-party protocols can be integrated with blockchain to further improve services recommendation systems. Combining credits of privacy-preservation of blockchain, Li et al.~\cite{li2019blockchain} proposed a blockchain-based QoS-aware web services recommendation framework. At the same time, Cai et al. ~\cite{cai2019personalized} proposed a personalized blockchain-based prediction approach. Both of them are the variants of matrix factorization. 

In summary, to guarantee the reliability of services users is a challenging task for services recommenders. Blockchain-based services recommenders can effectively address this issue while protecting the privacy of users. However, researches on blockchain-based services recommendation have yet been at a preliminary stage. For example, most studies only conduct insufficient experiments in a simplified network environment.

\begin{figure*}[t]
    \centering
    \includegraphics[width = 1\textwidth]{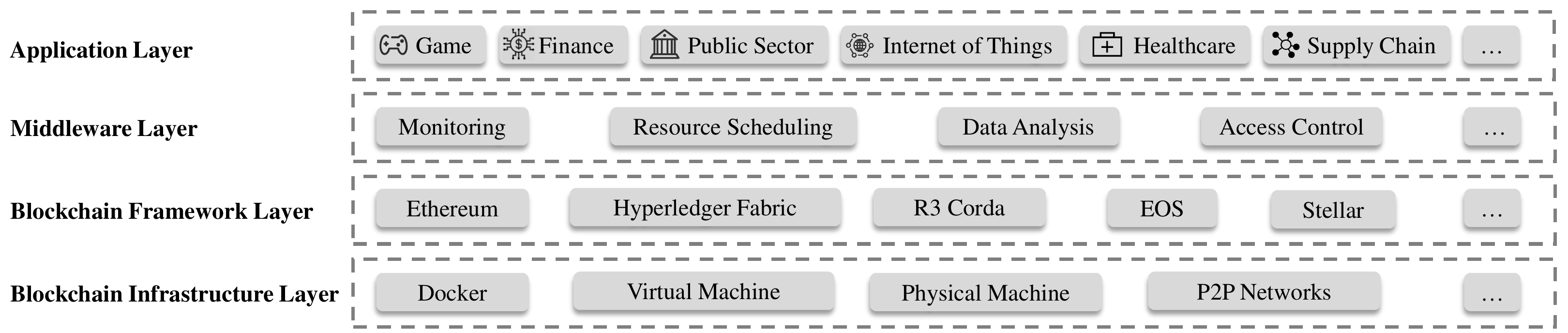}
    \caption{BaaS architecture}
    \label{fig:3}
\end{figure*}

\subsection{Services Arbitration Based on Blockchain}
Cloud services have recently been adopted by an increasing number of business institutions thanks to their high agility and superior cost performance index (CPI). However, a critical issue arises: no trusted-third-party ensures the correct execution of the workflow in a trustless environment. For example, services providers that do not obey the agreement may breach their obligations. The integration of service arbitration with blockchain technologies can potentially overcome the challenge.

There are a few studies on this issue. Zou et al.~\cite{zou2016dispute} present a services contract management scheme based on blockchain technology to address the issue of an untrusted distributed environment when executing services in a distributed network. In such a trusted environment, every participant that maintains a local services registry acts as an independent node in blockchain. This services contract management scheme eliminates the information asymmetry and information silo in services computing.

\section{Blockchain Based on Services Computing}
\label{sec:bconsc}
Up to now, blockchain has been proposed for more than a decade, but the following difficulties still hinder the impeding momentum of its progress. 
i) Scarcity of experienced developers. It is not easy to fully understand the working of such a composite blockchain system. Thus, developers usually have to spend a lot of time and effort in comprehending and controlling such a system, consequently leading to the difficulty in training experienced developers.
ii) Complex deployment and usage. We usually build a complex blockchain system for better scalability. Unfortunately, the increasing complexity of blockchain systems brings about a more complex deployment and usage of blockchain. 
iii) Costly maintenance. Ranging from a small application system to a large-scale distributed system, maintenance is always the most tiring, troublesome, and technical work. Particularly in blockchain, inherited from the distributed system, small network latency may be accumulated to services disruption. 

In this section, we first discuss how services computing affect blockchain and introduce the paramount application, \emph{Blockchain as a Service}, in Section~\ref{subsec:baas}. Next, we present BaaS architecture in Section~\ref{subsec:BaaS-arch} and BaaS platforms in Section~\ref{subsec:baas_plat}.


\subsection{Blockchain as a Service}
\label{subsec:baas}
To date, services computing, devoting to advocating the thought of encapsulating functions into services, is serving in various industries, and blockchain is no exception. We can thereby observe from the aforementioned challenges that a good choice to simplify the complex system is to provide a high-level abstraction of functionalities. There are various manners of integrating functions as a service (namely X as a Service), such as Infrastructure as a Service, Platform as a Service, Software as a Service. X as a Service manner can potentially enhance the scalability and reusability of blockchain. 

Therefore, \emph{BaaS} has been proposed to offer the services for blockchain to build and deploy, execute, monitor, and manipulate the business logic procedures across the entire enterprise. Deploying BaaS on cloud computing platforms, customers are allowed to leverage cloud-based solutions to design, develop, and host blockchain applications that can run on top of smart contracts and functions over blockchain networks. Services vendors here can manage all the necessary tasks and activities in order to support agile and operational infrastructures while customers only need to employ BaaS to execute tasks and activities~\cite{baasconcept}.

Moreover, abstracting the blockchain system into coarse-grained services within BaaS can help to construct decoupled services infrastructure. For example, the module that monitoring blockchain performance can be abstracted into a separate service, which is developed and maintained by a small group of specialists.
After further subdividing functionalities into services, the blockchain system offers a flexible combination of services, and each of them is relatively independent. Hence, by using these services, it brings convenience for developers to build a more complex blockchain-based application. 

Existing BaaS platforms have generally been implemented in a four-layer architecture (from top to bottom): \textit{Application Layer, Middleware Layer, Blockchain Framework Layer, Blockchain Infrastructure Layer} as shown in Fig.~\ref{fig:3}. BaaS architecture is further illustrated in Section~\ref{subsec:BaaS-arch}. In the past few years, tech giants established their own BaaS platforms including but not limited to Microsoft Azure, IBM Blockchain Platform, AWS. Other startups have also released their novel BaaS platforms. We summarize the representative BaaS platforms in Section~\ref{subsec:baas_plat}. 


\subsection{BaaS architecture}
\label{subsec:BaaS-arch}
BaaS creates a convenient, scalable, and high-performance blockchain ecosystem to develop, build, deploy, and manipulate blockchain applications. Current BaaS solutions can be usually divided into four layers (from top to bottom): \textit{Application Layer, Middleware Layer, Blockchain Framework Layer, Blockchain Infrastructure Layer}, as shown in Fig. \ref{fig:3}. Each layer can be treated as a single modular service that contains multiple simple functional components. We next introduce each layer in details:\\

\textit{{Application Layer}}. On top of the BaaS architecture is the application layer. Similar to other systems, blockchain-based applications are developed over the underlying blockchain infrastructure. In this manner, BaaS developers are not required to fully understand blockchain internals but only methods of invoking BaaS services to build a blockchain application. In this way, developers can focus on the business logic of applications while BaaS vendors only need to deploy decentralized applications on top of a blockchain infrastructure. As presented in Fig. \ref{fig:3}, BaaS supports the large-scale potential blockchain markets including but not limited to games~\cite{cryptokitties}, finance~\cite{mainelli2016chain}~\cite{cant2016smart}~\cite{de2015blockchain}, public sector~\cite{hyvarinen2017blockchain,olnes2016beyond}, IoT~\cite{chen2018iot,christidis2016blockchains}, healthcare~\cite{mettler2016blockchain,azaria2016medrec,hndai:IOTJM20} and supply chain~\cite{hofmann2017supply,korpela2017digital}.

\textit{{Middleware Layer}}. The second layer in the BaaS structure is the middleware layer, which acts as an interconnection agent between the application layer and the blockchain framework layer. On the one hand, the middleware layer hides the complexity of the underlying blockchain and offers the user-friendly interfaces to application developers in the above application layer. This layer essentially includes some fundamental system manipulation services such as monitoring, data analysis, resource scheduling, and access control. In particular, monitoring services ~\cite{zheng2018detailed} audit and measure the performance of blockchain systems and inform developers to take some actions when anomalies occur. Additionally, data stored on the public ledger can be collected and analyzed to develop blockchain applications~\cite{chen2018detecting}. Resource scheduling algorithms approach the optimal resource allocation to reach a maximal utility of limited resources. Since access control is to regulate the access of data, data security can be guaranteed in BaaS with appropriate authentication and authorization. As a result, the middleware layer guarantees the reliability and scalability of BaaS.

\textit{{Blockchain Framework Layer}}. The third layer is to construct a blockchain framework based on blockchain infrastructure. The blockchain framework generally refers to the smart contract platform as mentioned in Section \ref{sec:background}. In this layer, a smart contract platform provides tokenized programs written by specific languages and specified by execution environments (e.g., virtual machines). From the perspective of stability and diversity, the BaaS framework supports popular smart contract platforms such as Hyperledger Fabric, Ethereum, Quorum (Ethereum for enterprise), R3 Corda, EOS, Stellar at present. Blockchain developers can therefore build the business logic written in specific program codes running on top of stable environments. 

\textit{{Blockchain Infrastructure Layer}}. The blockchain infrastructure layer locates at the bottom of BaaS architecture. Since blockchain is essentially a distributed system, blockchain infrastructure offers communication/networking services with corresponding computational resources such as the physical machines, virtual machines, or the emerging docker containers to execute smart contracts. Both the difficulty to deploy blockchain infrastructure and the high cost of blockchain operation and maintenance impede the development momentum in the blockchain. Fortunately, the blockchain infrastructure layer deployed on top of existing cloud platforms thereby allows developers to build their blockchain applications without the necessity of establishing underneath networks from scratch.

\subsection{Current BaaS Platforms}
\label{subsec:baas_plat}
\begin{table*}
\renewcommand\arraystretch{2}
\caption{Comparison of Blockchain as a Service (BaaS) platforms}
\begin{center}
\begin{tabular}{m{3.5cm}<{\centering}|m{4cm}<{\centering}cm{4cm}<{\centering}cm{4cm}<{\centering}}
\hline
\textbf{BaaS Platform}& \textbf{Azure Blockchain Workbench} & \textbf{IBM Blockchain Platform}& \textbf{AWS Blockchain Platform}   \\
\hline\hline
\textbf{Blockchain Infrastructure} & Azure & Kubernetes & ECS, EC2 \\
\hline

\textbf{Blockchain Framework} &  Ethereum, Hyperledger Fabric, Quorum and R3 Corda & Hyperledger Fabric & Ethereum, Hyperledger Fabric \\
\hline
\textbf{Supporting Services}& REST APIs for managing and message-based APIs for integration & Dev tools and Operation tools  & Managed Services and Amazon QLDB \\
\hline
\textbf{Target User} & Mostly for enterprise use & For various enterprise use & For retail customers and multi-parties  \\

\hline
\end{tabular}
\label{tab2}
\end{center}
\end{table*}


With the advances of blockchain technologies, many tech giants (such as Microsoft, IBM, Amazon) established their own BaaS platforms on top of existing cloud computing infrastructures. Most of these BaaS platforms have been developed on the four-layer architecture as introduced in Section~\ref{subsec:BaaS-arch}. In this section, we will briefly introduce representative BaaS platforms of IBM, Microsoft Azure, and AWS from the perspective of the four-layer architecture and make a clear comparison between them from perspectives of blockchain infrastructure, blockchain framework, supporting services, target users as summarized in Table \ref{tab2}. \\

\subsubsection{\textbf{Azure Blockchain Workbench}}
Azure Blockchain Workbench~\cite{azure} released by Microsoft is one of the first major BaaS platforms to companies and developers. It claims to provide a low-risk, low-cost, fast, and fault-tolerance blockchain solution to business in the BaaS manner. Azure Blockchain Workbench essentially constructs the BaaS platform using the existing cloud computing platform \textit{Microsofe Azure} as the infrastructure for the blockchain infrastructure layer. To improve its scalability, Azure Blockchain Workbench supports several mainstream blockchain frameworks like Ethereum, Hyperledger Fabric, and R3 Corda. Also, Azure Blockchain Workbench provides RESTful APIs to manage blockchain applications and users, and message-based APIs to integrate BaaS with existing systems.


\subsubsection{\textbf{IBM Blockchain Platform}} IBP~\cite{ibm} can automate commercial activities through digitizing working flows (in a form of transactions) over the distributed ledger to ensure security, inter-operation, and reliability. IBP~\cite{ibmwhitepaper} is running on top of Kubernetes architecture, which is serving as the infrastructure in BaaS architecture. Meanwhile, IBP leverages Hyperledger Fabric as the blockchain framework with the provision of high security, scalability, and quick response, which are crucial to modern businesses today. In the Middleware layer, IBP provides lots of value-added tools including dev tools and operation tools to ensure the scalability, flexibility, governance over the blockchain network. The main objective of IBP is to build a BaaS platform for various enterprises so that IBP offers four kinds of services including Starter Plan, Enterprise Plan, Enterprise Plus Plan, and Remote Peer.

\subsubsection{\textbf{Amazon Web Services}}

AWS~\cite{aws,awswhitepaper}, the cloud computing business operated by e-commerce giant Amazon, released the Amazon blockchain platform in April 2018. AWS blockchain platform consists of three parts: AWS Blockchain Templates, AWS Managed Blockchain, and Amazon Quantum Ledger Database (QLDB). AWS Blockchain Template deploys the open-source blockchain framework, which is chosen by developers as containers on an Amazon Elastic Container Service (ECS) cluster or directly on an Amazon Elastic Compute Cloud (EC2) instance running atop Docker. Amazon Managed Blockchain can support creating and managing scalable blockchain networks such as Hyperledger Fabric and Ethereum, which acts as the middleware on BaaS. Additionally, AWS Managed Blockchain can replicate each network operation into Amazon QLDB. In this way, network activities can be analyzed and visualized. In short, AWS provides blockchain solutions for retail customers and multi-parties. 






\section{Future Directions}
\begin{figure}[t]
    \centering
    \subfigure[Open issues in Blockchain-based Services computing]{\label{fig:openissuesinbbsc}\includegraphics[width = 0.45\textwidth]{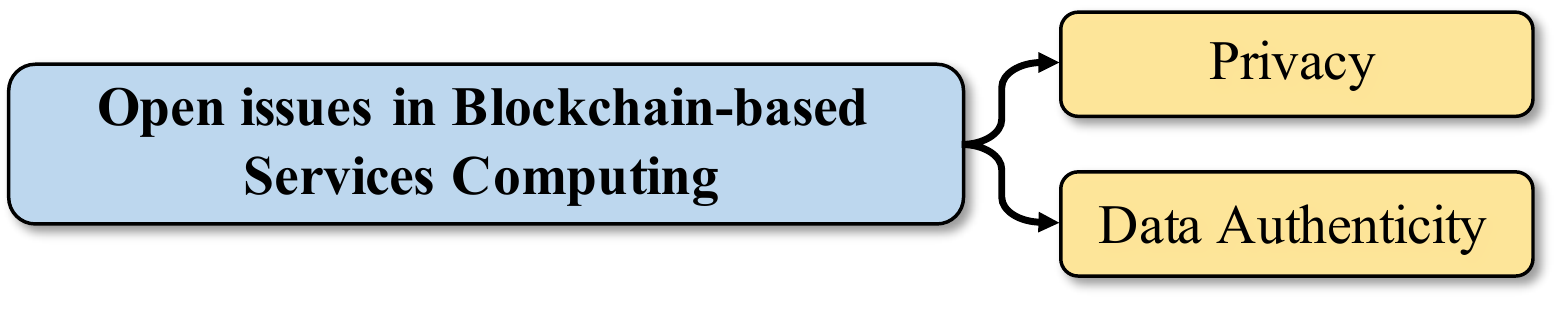}    }
    \subfigure[Open issues in BaaS]{\label{fig:openissuesinbaas}\includegraphics[width = 0.45\textwidth]{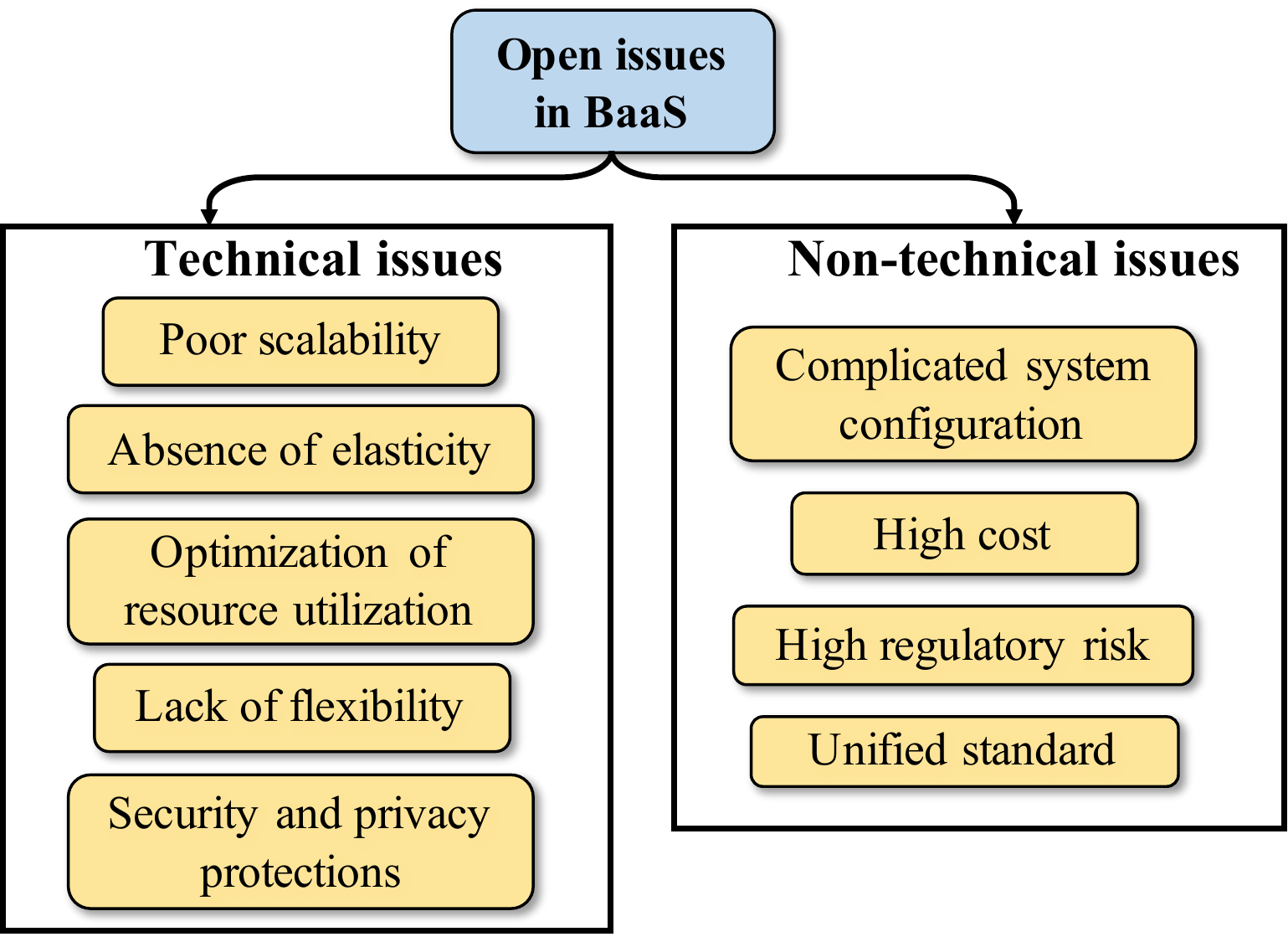}}     
    \caption{Overview of future directions}
    \label{fig:openissues}
\end{figure}

Despite opportunities brought by the integration of blockchain with services computing, there are still scores of open issues such as privacy and data authenticity in blockchain-based services computing as discussed in Section~\ref{subsec:challengesOfbbsc}. Meanwhile, as an emerging area, BaaS still faces many challenges exhibiting in technical and non-technical aspects, which will be discussed in Section~\ref{subsec:challenges}. Fig. \ref{fig:openissuesinbaas} outlines the open issues in blockchain-based services computing and Fig. \ref{fig:openissuesinbaas} gives an overview of the open issues in BaaS. 

\subsection{Open issues in Blockchain-based Services Computing}
\label{subsec:challengesOfbbsc}
Although blockchain brings many opportunities to address existing issues in services computing, there are a number of open issues to be resolved in aspects of privacy, and data authenticity, which will be discussed in detail as follows.

\subsubsection{\textbf{Privacy}}
Although blockchain can offer pseudonymity of blockchain users, which are anonymous in the interactions of blockchain via their public addresses, on-chain data in blockchain are essentially available to be accessible for everyone. Thus, how to protect blockchain data privacy while ensuring effective data sharing especially in services computing is a critical issue.

Recently, several potential solutions have been proposed to preserve blockchain data privacy. These schemes include mixed coins, homomorphic encryption, zero-knowledge proof, ring signature. In particular, Zhao et al.~\cite{zhao2019machine} proposed a method to integrate multiple schemes such as ring signature and double-authentication-preventing signature to ensure the security while adopting a similarity learning method to guarantee the availability of trading data and consequently protect the privacy of data providers. Gai et al.~\cite{gai2019privacy} proposed an approach to mainly protect the privacy of energy trading users in smart grid and to inspect the distribution of energy selling.

\subsubsection{\textbf{Data Authenticity}}
Data authenticity is also a crucial issue in services computing. Although smart contracts can somehow guarantee the blockchain data authenticity, they cannot guarantee the authenticity of off-chain data (generated from external services).


To address this challenge, an additional agent, called Oracle is designed for data exchanging across the chain. The Oracle acts as a data carrier to mediate invocation from smart contracts and external services within the inter-organizational business processes. Nowadays, several proposals have been put forth based on Oracle. Oraclize~\cite{oraclize} is one of the successful oracle vendors to serve as a notary to fetch data directly from the smart contract requires. However, the authentication in Oracle has not been well addressed yet.

\subsection{Open issues in BaaS}
\label{subsec:challenges}

Services vendors are attempting to develop robust and scalable BaaS platforms yet the development of BaaS is at a very preliminary stage. Many provisioned services are not suitable to be deployed in practical industrial environments. Such issues are expected to be resolved in the future. We categorize these issues into two types: technical issues and non-technical issues, which are discussed as follows.


\subsubsection{\textbf{Technical issues}} Integrating with blockchain technologies, technical issues also cover the whole life cycle of BaaS. Thus each component inside BaaS has a more or less impact concerning the effectiveness of BaaS. We simply conclude these technical difficulties of each component in BaaS from a high-level abstraction. 
\begin{itemize}
    \item \textit{Poor scalability}. Existing blockchain frameworks can only process a small number of on-chain transactions in order to support small scale applications such as the public sector and healthcare, which have low requirements on the system throughput, i.e., less than 10 transactions per second (TPS). However, they cannot cater to the increasing demands from large scale applications such as financial and IoT services, which may require several thousand TPS.

    \item \textit{Absence of elasticity}. Many business applications may have dynamic requirements for system performance. For example, an application may request a huge amount of resources at a certain time while its requirement may significantly affect the QoS of other users. However, existing blockchain systems cannot fulfill the burst demands elastically. It is necessary to monitor the realtime performance and tune blockchain systems to fulfill the dynamic demands.

    \item \textit{Optimization of resource utilization}. It is often less cost-effective to purchase and deploy extract computing facilities to meet the elasticity and scalability requirements. Thus, it is necessary to optimize the resource usage via maximizing the spare computational resources and improving the total utilization of various components of the system. It is worthwhile to investigate several open issues: 1) the quantitative analysis of resource utilization, 2) optimization strategies (e.g., incentive, pricing, and optimization mechanisms) to maximize the resource utilization.

    \item \textit{Lack of flexibility}. Current BaaS platforms cannot fulfill various application scenarios. Future BaaS should be pluggable to diverse application scenarios. In particular, the serverless architecture has gained much attention in the past few years and it is expected to be mainstream architecture for future services computing. In the serverless architecture, an application will be resolved into diverse independent components that provide general interfaces like APIs to adapt to different systems~\cite{chen2018fbaas}. When blockchain is integrated into serverless architecture in the future, users will be able to customize their applications based on diverse underlying blockchain systems. However, some technical issues are expected to be addressed before the fusion of blockchain with a serverless architecture.

    \item \textit{Security and privacy protections}. On the one hand, blockchain systems also have their own security vulnerabilities such as border gateway protocol (BGP) routing hijack attack and decentralized autonomous organization (DAO) attack in smart contracts~\cite{ZZheng:FGCS20}. On the other hand, data stored on the public ledger can be visible to everyone despite the user pseudonymity of blockchain. It is shown in~\cite{hndai:blockchain-iot2019} that extensive analysis of multiple transactions can essentially identify one common user account. Therefore, security assurance and privacy protection are still open issues in BaaS research community.
\end{itemize}

\subsubsection{\textbf{Non-technical issues}}Apart from technical issues, non-technical issues also hinder the development of BaaS. We describe several typical non-technical issues as follows.

\begin{itemize}
    \item \textit{Complicated system configuration}. Although most of BaaS solutions provide a whole life cycle of blockchain services, they generally only offer services interfaces for creating blockchain applications. The complexity of underlying blockchain infrastructure and various frameworks also exhausts developers who need to learn extensive blockchain knowledge (such as cryptography and decentralized consensus) from scratch.
    \item \textit{High cost}. The incumbent blockchain systems are suffering from the high equipment cost and maintenance expenditure, which reduce the motivations of enterprises to adopt BaaS solutions. Cost-effective  BaaS solutions are expected to be further explored in the future.
    \item \textit{High regulatory risk}. Even though many countries encourage business institutes to transform conventional business modes into blockchain-based modes, the process of transforming the information chain to the value chain also brings financial risks as well as regulatory difficulty. For example, it is difficult to monitor and prevent the decentralized blockchain from malicious behaviours (such as money laundering). Thus, it is also an open question to design blockchains that are regulatory while still maintaining decentralization. 
    
    \item \textit{Unified Standard}. The absence of a unified standard for diverse blockchain systems is a root cause for many technical issues (e.g., the absence of elasticity and flexibility). Similar to the development of the Internet, which has defined a general protocol stack to unify all message formats and interaction manners, we believe that the future blockchain community will also form a unified set of rules or modular protocols/schemes to support diverse blockchain systems. The unified blockchain standard will help to overcome the obstacles across different blockchain systems and improve the application flexibility, consequently promoting the development of BaaS. 

\end{itemize}

\section{Conclusion}
\label{sec:conc}
This paper firstly presents an overview on blockchain and services computing. We investigate the state-of-the-art literature on services computing and summarize three major concerns: i) pricing mechanisms and incentives, ii) information silo, iii) security and privacy risks. The four key characteristics of the emerging blockchain technology can potentially address the above challenges of services computing. We then analyze services computing based on blockchain from five perspectives: services creation, services discovery, services composition, services recommendation, services arbitration. Furthermore, we briefly review existing BaaS solutions and compare them from the hierarchical architecture of BaaS. We also summarize key issues both in blockchain-based services computing and BaaS. In summary, we believe that the advent of blockchain technology will promote the renaissance of services computing.


\balance

\bibliography{Blockchain_ref}

\end{document}